\documentclass[conference]{IEEEtran}
\usepackage{graphicx}
\usepackage{amsmath}
\usepackage{subfigure}
\usepackage{url}
\usepackage{color}
\usepackage{geometry}
\usepackage{graphicx}
\usepackage{amssymb}
\usepackage{epstopdf}
\usepackage{amsmath}
\usepackage[ruled,vlined,linesnumbered]{algorithm2e}
\newtheorem{theorem}{\textbf{Theorem}}
\newtheorem{corollary}{\textbf{Corollary}}

\ifCLASSINFOpdf
\else
\fi
\IEEEoverridecommandlockouts
\begin{document}
\title{The Urge to Merge: When Cellular Service Providers Pool Capacity}

\author{\IEEEauthorblockN{Sha Hua, Pei Liu and Shivendra S. Panwar$^*$ \thanks{*The authors are with Polytechnic Institute of New York University, Brooklyn, NY, 11201, USA. Emails: shua01@students.poly.edu; pliu@poly.edu; panwar@catt.poly.edu}
}
%\IEEEauthorblockA{Polytechnic Institute of New York University, Brooklyn, NY, 11201, USA}
%\IEEEauthorblockA{shua01@students.poly.edu; pliu@poly.edu; panwar@catt.poly.edu}
}


%Telephone: (800) 555--1212, Fax: (888) 555--1212}


\maketitle

\begin{abstract}
%\boldmath
As cellular networks are turning into a platform for ubiquitous data access, cellular operators are facing a severe data capacity crisis due to the exponential growth of traffic generated by mobile users. In this work, we investigate the benefits of sharing infrastructure and spectrum among two cellular operators. Specifically, we provide a multi-cell analytical model using stochastic geometry to identify the performance gain under different sharing strategies, which gives tractable and accurate results. To validate the performance using a realistic setting, we conduct extensive simulations for a multi-cell OFDMA system using real base station locations. Both analytical and simulation results show that even a simple cooperation strategy between two similar operators, where they share spectrum and base stations, roughly quadruples capacity as compared to the capacity of a single operator. This is equivalent to doubling the capacity per customer, providing a strong incentive for operators to cooperate, if not actually merge.
\end{abstract}
\vspace{-0.1in}


% \fi
\IEEEpeerreviewmaketitle

\section{Introduction}

The recent commercialization and popularization of 3G networks have significantly enhanced the mobile users' capability of making use of ubiquitous data access. However, as users ``mobilize'' all of their communication activities, from streaming video to cloud computing tasks, cellular operators are facing a severe data capacity crisis. A recent study of Cisco predicts that mobile data traffic will skyrocket by a factor between 25X to 50X by 2015~\cite{CiscoForecast}. Such explosive data traffic growth will overload the cellular infrastructure and result in either poor or expensive service to the subscribers. To address this challenge, the most straightforward solutions, such as adding cells and increasing spectrum have become either expensive or inefficient options.

Cooperation among operators is yet another solution which has not drawn much attention from researchers. As cellular operators are devoting most of efforts in expanding their respective networks due to the urgency of the data capacity crisis and market competition, it has become apparent that there exist huge variations in spectrum usage, channel quality and coverage in different operators' networks~\cite{CoverageData}. Such diversity generates plenty of cooperation opportunities, which can be exploited to improve their network performance, spectrum efficiency and user experience.

\begin{figure}[htbp]
    \centerline{\includegraphics[scale=0.75]{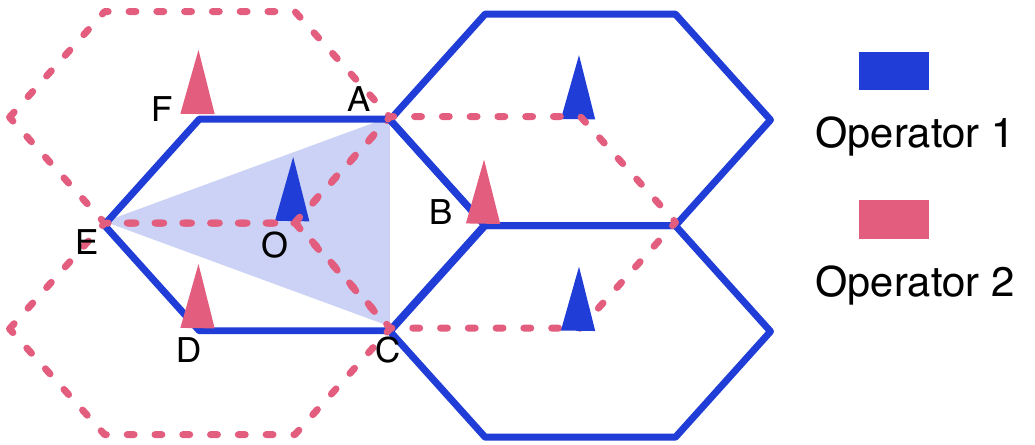}}
    \caption{
       Motivating scenario of cooperation between cellular operators.
    }
    \label{fig:motivatingexample}
\vspace{-0.25in}
\end{figure}
One motivating example is shown in Figure~\ref{fig:motivatingexample}. Two operators, Operator 1 and Operator 2 (OP$_1$ and OP$_2$ for short), provide cellular service in one area. We assume their base stations (BSs) are deployed following a hexagonal layout (solid and dotted lines, respectively) and are interweaved. Without cooperation, the two operators will serve their own users solely using their own spectrum bands. The edge users, such as those at points B, D and F in the left cell of OP$_1$, will experience bad channel condition and strong inter-cell interference. However, if the two operators cooperate by allowing their users to flexibly connect to any BS that is closest to them, users of OP$_1$ at B, D and F will be served by the BSs of OP$_2$ and have excellent channels to the BSs. Generally, the users of OP$_1$ in the area $\triangle ABC$, $\triangle ECD$ and $\triangle EFA$ will enjoy performance gains; these are mostly edge users with low data rates. A similar effect happens to the users of OP$_2$ as well, such as those in the area $\triangle AOC$. As we can see, simple cooperation improves the capacity of both operators and results in a win-win situation. Depending on the levels of synergy, the operators can share each other's infrastructure, spectrum, or even merge into a single operator. Interestingly, in AT\&T's proposed acquisition of T-Mobile, one of the reasons cited is the resulting increase in capacity~\cite{FCCMerger11}.

In this paper, we focus on studying the benefits brought by different levels of cooperation among cellular operators. We are interested in answering the following questions: In what ways can the operators cooperate? How much performance improvement can their cooperation bring? Only a few recent papers address this issue~\cite{Singh@TON11}\cite{Supratim@MOBICOM11}\cite{Peng@WC10}. In~\cite{Singh@TON11}, Chandramani et al. optimize the operators' aggregate payoffs given BS locations and user rates, and use game theory to discuss how to share the profits. Supratim et al. in~\cite{Supratim@MOBICOM11} develop a user choice algorithm with network information provided by the operators. Peng et al. in~\cite{Peng@WC10} focus on spectrum-based cooperation, and form a group bargaining model based on the demand on each BS.

However, all the above papers require cross-layer optimization based on the instantaneous network status. They do not characterize the long-term network capacity improvement for a large-scale multi-cell deployment. Our research fills this gap and provides a tractable and accurate model for quantifying the gains under different cooperation strategies. To validate our model, we have collected a set of real BS location data and provide performance results using real-world layouts under an Orthogonal Frequency Division Multiple Access (OFDMA) system. This aspect also has not yet been examined before. We expect our paper to give an insight into the benefit of cooperation and guide operators considering a range of cooperation options. Specifically the contribution of the paper is two-fold:
\begin{itemize}
\item We present an analytical model for identifying the performance gain from cooperation among cellular operators. It provides a tractable and reasonably accurate model for average user rate/throughput under a multi-cell environment.
\item We perform extensive simulations using real BS locations and OFDMA resource allocation algorithms. It validates the advantage of operator cooperation in a realistic setting.
\end{itemize}

The remainder of the paper is organized as follows. Section II presents two modes of cooperation and the analytical model. In Section III, we provide an OFDMA resource allocation algorithm. The performance evaluation is presented in Section IV. Finally, the paper is concluded in Section V.

\section{Analytical Modeling}

In this section, we derive an analytical model to evaluate the network performance when the cellular operators cooperate. We consider two strategies, with different levels of cooperation, that we envisage the cellular operators may adopt:
\\ \indent 1) \textbf{FLEXROAM} (short for ``flexible roaming''): the cellular operators allow their users to freely connect to a BS of either operator that provides the best signal strength. It will require an update in the signaling protocols to facilitate this. \\
\indent 2) \textbf{MERGER}: in addition to FLEXROAM, all the BSs of an operator can reuse the spectrum of its cooperating operators. This may be the result of a full merger of the two operators, but does not preclude a mutual business agreement short of a full merger.
%\begin{itemize}
%\item \textbf{FLEXROAM} (short for ``flexible roaming''): the cellular operators allow their users to freely connect to BS of either operator that provides the best signal strength. It will require an update in the signaling protocols to facilitate this.
%\item \textbf{MERGER}: in addition to FLEXROAM, all the BSs of an operator can reuse the spectrum of its cooperating operators. This refers to an additional sharing of spectrum, which may be the result of a full merger of the two operators, but does not preclude a mutual business agreement short of a full merger.
%\end{itemize}

Note that FLEXROAM and MERGER can also be offered to users by a mobile virtual network operator, or MVNO, which can purchase wholesale use of the infrastructure from the two operators. We further use \textbf{NOCOOP} to refer to the scheme that the two operators do not cooperate.

Cellular networks are traditionally modeled by assuming the BSs are placed following a hexagonal layout~\cite{Theodore@BOOK02}. However, these models have long suffered from being both intractable and highly idealized. Recently, a general model based on stochastic geometry was proposed in~\cite{Jeffrey@TCOM10}, which provides tractable ways to evaluate network performance considering inter-cell interference and fading. Our analytical modeling follows the methods in~\cite{Jeffrey@TCOM10}, and provides tractable results of the performance under the various modes of operators' cooperation.

We will mainly focus on the user's \textit{average ergodic rate} and \textit{average throughput}, which are important metrics for benchmarking the cellular system. We assume there are two operators, OP$_1$ and OP$_2$, coexisting in the network. Our models can be easily extended to include more than two operators, and to derive other performance metrics, such as the distribution of user's Signal-to-Interference-plus-Noise Ratio (SINR).

We assume the locations of BSs of OP$_1$ and OP$_2$ follow independent Poisson point processes $\Phi_1$ and $\Phi_2$ with densities $\lambda_1$ and $\lambda_2$, respectively. The model drawn from such a random deployment is shown to be about as accurate as the standard grid model, compared to an actual cellular network~\cite{Jeffrey@TCOM10}. Further we use $W_i$ to denote channel bandwidths used by the BSs of OP$_i$, $i \in \{1, 2\}$. We consider the downlink performance of the system. Without losing generality, we assume a typical user is located at the origin. If it is connected to the BS $b_0$ of OP$_i$, its SINR can be expressed as:
\begin{equation}
\textrm{SINR}_{i} = \frac{P_t h_{b_0} r_{b_0}^{-\alpha}}{N_0 W_i + P_t \sum_{b \in \Phi_i \setminus b_0}g_{b} r_{b}^{-\alpha}},
\end{equation}
where $P_t$ is the fixed transmission power of all the BSs, $N_0$ is the noise power density. The distance between the user and BS $b_0$ is $r_{b_0}$, and the channel fading is $h_{b_0}$. The user's distances to other interfering BSs $b$ in $\Phi_i$ are $r_{b}$ and the corresponding channel fadings are $g_{b}$. $\alpha$ is the path loss exponent. We assume all the fadings are Rayleigh fading with mean 1.

\subsection{Average User Ergodic Rate}

Following the proofs in~\cite{Jeffrey@TCOM10}, when the two operators do not cooperate and run their service as is, the average ergodic rate in the downlink for a typical user of OP$_i$'s is given by:
\begin{eqnarray} \label{eqn:2}
&& R_{NOCOOP}(W_i, \lambda_i) = \mathbb{E}[\ln (1 + \textrm{SINR}_i)]  \nonumber \\
&=& \int_{r>0}e^{-\pi \lambda_i r^2} \int_{t>0}e^{-\frac{N_0 W_i r^\alpha (e^t - 1)}{P_t}} F(\lambda_i) \mbox{d}t 2\pi \lambda_i r \mbox{d}r,
\end{eqnarray}
where
\begin{equation} \label{eqn:3}
F(\lambda_i) = \exp(-\pi \lambda_i r^2(e^t - 1)^{2/\alpha} \int_{(e^t - 1)^{-2/\alpha}}^{\infty} \frac{1}{1 + x^{\alpha / 2}} \mbox{d}x).
\end{equation}

Eq.~(\ref{eqn:2}) and~(\ref{eqn:3}) can be computed numerically and are used for performance comparisons in this paper. Next we examine the performance under cooperation strategies. When OP$_1$ and OP$_2$ share infrastructure following the FLEXROAM strategy, we have the following theorem:
\begin{theorem}
Under FLEXROAM strategy, for a typical user, the average ergodic rate is given by:
\begin{eqnarray}
&&R_{FLEXROAM}(\mathbf{W}, \boldsymbol \lambda) \nonumber \\
&=& \frac{\lambda_1}{\lambda_1 + \lambda2}R_{1}(W_1, \boldsymbol \lambda) + \frac{\lambda_2}{\lambda_1 + \lambda2}R_{2}(W_2, \boldsymbol \lambda),
\end{eqnarray}
where
\begin{eqnarray}
R_{i}(W_i, \boldsymbol \lambda) &=& \int_{r>0}e^{-\pi (\lambda_1 + \lambda_2) r^2} \int_{t>0}e^{-\frac{N_0 W_i r^\alpha (e^t - 1)}{P_t}} \cdot \nonumber \\
&&F(\lambda_i) \mbox{d}t 2\pi (\lambda_1 + \lambda_2) r \mbox{d}r.
\end{eqnarray}
and $F(\lambda_i)$ is given by Eq. (3). $\mathbf{W} = \{W_1, W_2\}$ and $\boldsymbol \lambda = \{\lambda_1, \lambda_2\}$ are the vectors for bandwidths and BS densities.
\end{theorem}
\begin{IEEEproof}
We use $r$ to denote the distance between the user and its associated BS. Under FLEXROAM, the user will connect to the closest BS from either of the operators that gives the best average signal strength. The union of all the BSs is still a Poisson point process $\Phi$ with density $\lambda_1 + \lambda_2$. The cumulative density function (CDF) of the distance $r$ is then:
\begin{eqnarray}
F_r(R) &=& \textrm{Pr}[r \leq R] = 1 - \textrm{Pr}[r > R] \nonumber \\
&=& 1 - \textrm{Pr}[\textrm{no BS from any operator closer than $R$}] \nonumber \\
&=& 1 - e^{-(\lambda_1 + \lambda_2) \pi R^2}.
\end{eqnarray}
Thus the probability density function (PDF) of $r$ is:
\begin{equation}
f_r(r) = \frac{\mbox{d}F_r(r)}{\mbox{d}r} = e^{-(\lambda_1 + \lambda_2) \pi r^2} 2 \pi (\lambda_1 + \lambda_2) r.
\end{equation}
Moreover, we have
\begin{equation} \label{eqn:8}
R_{FLEXROAM}(\mathbf{W}, \boldsymbol \lambda) = \int_{r>0} f_r(r) \mathbb{E}_{h, g, \Phi}(\ln(1 + \textrm{SINR})) \mbox{d}r.
\end{equation}
We further use $X_i = 1$ to represent the event that the user is associated to the BS of OP$_i$. Note $X_i \in \{0, 1\}, \forall i = 1, 2$ and $X_1 + X_2 = 1$. According to the rule of total probability,
\begin{equation} \label{eqn:9}
\mathbb{E}_{h,g,\Phi}(\ln(1 + \textrm{SINR})) = \sum_{i = 1, 2}\mathbb{E}_{h,g,\Phi_i}(\ln(1 + \textrm{SINR}_i)) \textrm{Pr}[X_i = 1],
\end{equation}
wherein, SINR$_i$ is given by Eq. (1). Using $D_i$ to represent the user's distance to the closest BS from OP$_i$, in a manner similar to Eq.~(7), the CDF of $D_i$ is $f_{D_i}(r) = e^{-\lambda_i \pi r^2} 2 \pi \lambda r$. We have:
\begin{eqnarray} \label{eqn:10}
&& \textrm{Pr}[X_1 = 1] = \textrm{Pr}[D_1 < D_2] \nonumber \\
&=& \int_{0}^{\infty} \int_{r_1}^{\infty} e^{-\lambda_2 \pi r_2^2} 2 \pi \lambda_2 r_2 e^{-\lambda_1 \pi r_1^2} 2 \pi \lambda_1 r_1 \mbox{d}r_2 \mbox{d}r_1 \nonumber \\
&=& \lambda_1 / (\lambda_1 + \lambda_2).
\end{eqnarray}
Similarly, we have $\textrm{Pr}[X_2 = 1] = \lambda_2 / (\lambda_1 + \lambda_2)$. Plugging Eq.~(\ref{eqn:9}) and~(\ref{eqn:10}) into~(\ref{eqn:8}), and following the rest of the steps as in Appendix B of~\cite{Jeffrey@TCOM10} completes the proof.
\end{IEEEproof}

We next discuss the average ergodic rates when the MERGER strategy is employed. Under this scheme, each BS reuses the spectrum owned by the other operators, e.g., they can operate on the whole spectrum $W_1 + W_2$. As a result, the BSs from both operators interfere with each other. We then have the following corollaries from Eq. (2) and (3):
%\begin{corollary}
%Under the SPECTRUM SHARING strategy, for a typical user of OP$_i$, the average ergodic rate is:
%\begin{eqnarray}
%&&R_{SS_i}(\mathbf{W}, \boldsymbol \lambda) =  \int_{r>0}e^{-\pi \lambda_i r^2} \cdot \nonumber \\
%&&\int_{t>0}e^{\frac{P_t r^\alpha (e^t - 1)}{N_0 (W_1 + W_2)}} \cdot F(\lambda_1 + \lambda_2) \mbox{d}t 2\pi \lambda_i r \mbox{d}r
%\end{eqnarray}
%\end{corollary}
%\begin{IEEEproof}
%Since all the BSs of both operators are interfering and the bandwidth is increased to $W_1 + W_2$, the new SINR of a typical user is represented by replacing $W_i$ with $W_1 + W_2$ and $\Phi_i$ with $\Phi_1 \cup \Phi_2$ in Eq.~(1).
%%\begin{equation}
%%SINR_{new} = \frac{P_t h r^{-\alpha}}{N_0 (W_1 + W_2) + P_t \sum_{k \in \Phi_1 \cup \Phi_2 / b_0}g_k r_k^{-\alpha}},
%%\end{equation}
%Moreover, the pdf of the distance $r$ between the user in OP$_i$ and its BS is still $f_{r_i}(r) = e^{-\lambda_i \pi r^2} 2 \pi \lambda r$. Plug these into the proofs in Appendix B of~\cite{Jeffrey@TCOM10}, we derive the results.
%\end{IEEEproof}
\begin{corollary}
Under the MERGER strategy, for a typical user, the average ergodic rate is:
\begin{equation}
R_{MERGER}(\mathbf{W}, \boldsymbol \lambda) = R_{NOCOOP}(W_1 + W_2, \lambda_1 + \lambda_2).
\end{equation}
\end{corollary}
\begin{IEEEproof}
This is same as the single operator case with BSs of density $\lambda_1 + \lambda_2$ operate on bandwidth $W_1 + W_2$. We then have the corollary following Eq.~(2) and~(3).
\end{IEEEproof}

\subsection{Average User Throughput}

We will now determine the average throughput for a typical user under FLEXROAM and MERGER. We use $\eta_i$ to denote the subscriber density of OP$_i, i \in \{1,2\}$. Generally, subscriber density is far larger than the macrocell BS density, thus we assume $\eta_i \geq \lambda_j, \forall i, j  \in \{1,2\}$.

According to the Law of Large Numbers, in NOCOOP, if the two operators serve their users independently, the number of users in each cell of OP$_i$ will be approximately $\frac{\eta_i S}{\lambda_i S} = \frac{\eta_i}{\lambda_i}$, where $S$ is the total area under consideration. We further assume that the scheduling decisions on the BSs ensure \textit{proportional fairness}, which is a common design objective of all current and next generation cellular systems. With proportional fairness, the average throughput of a user from OP$_i$ is its rate divided by the number of users in the cell. Under NOCOOP scheme, it is:
\begin{equation}
Th_{NOCOOP}(W_i, \lambda_i, \eta_i) = R_{NOCOOP}(W_i, \lambda_i) \cdot \frac{ \lambda_i}{\eta_i},
\end{equation}
%\begin{equation}
%Th_{SS_i}(\mathbf{W}, \boldsymbol \lambda, \eta_i) = R_{SL_i}(\mathbf{W}, \boldsymbol \lambda)  \cdot \frac{ \lambda_i}{\eta_i}, \ i \in {1, 2}.
%\end{equation}

Moreover, under FLEXROAM and MERGER strategy, the number of users in each cell converges to $\frac{\eta_1 + \eta_2}{\lambda_1 + \lambda_2}$. Therefore, the average throughput for a typical user is:
\begin{eqnarray}
& Th_{FLEXROAM}(\mathbf{W}, \boldsymbol \lambda, \boldsymbol \eta) = R_{FLEXROAM}(\mathbf{W}, \boldsymbol \lambda) \frac{\lambda_1 + \lambda_2}{\eta_1 + \eta_2} \nonumber \\
&= \frac{1}{\eta_1 + \eta_2} (\lambda_1 R_1(\lambda_1, \lambda_2) + \lambda_2 R_2(\lambda_1, \lambda_2)).
\end{eqnarray}
where $\boldsymbol \eta = \{{\eta_1, \eta_2}\}$. Finally, under the MERGER strategy, the average throughput for a user is:
\begin{equation}
Th_{MERGER}(\mathbf{W}, \boldsymbol \lambda, \boldsymbol \eta) = R_{MERGER}(\mathbf{W}, \boldsymbol \lambda) \frac{\lambda_1 + \lambda_2}{\eta_1 + \eta_2}.
\end{equation}

Compared to the traditional hexagonal model that requires complex system-level Monte-Carlo simulations, our analytical model can be computed using a numerical tool in a short time.

\section{An OFDMA-based Multi-Cell System}

In the previous section, we described an analytical model for evaluating the network performance under different cooperation strategies of the operators. However, the model simplifies the real system in three aspects: 1) it assumes Poisson random BS deployment; 2) it uses the entire spectrum without considering subchannelization; 3) it assumes perfect resource allocation following the proportional fairness criterion.

Due to the above constraints, it is still good to validate what the network performance will be in a practical multi-cell system under the various cooperation strategies. To evaluate the benefits of operator cooperation in realistic system, we plan to combine real BS location data and practical user-subchannel scheduling algorithms. As OFDMA is widely adopted by all the next-generation cellular standards (WiMAX, LTE) to divide the spectrum into subchannels, in this section, we present a low complexity OFDMA multi-cell resource allocation algorithm. Results with real BS locations are given in the next section.

\begin{figure}[htbp]
    \vspace{-0.1in}
    \centerline{\includegraphics[scale=0.3]{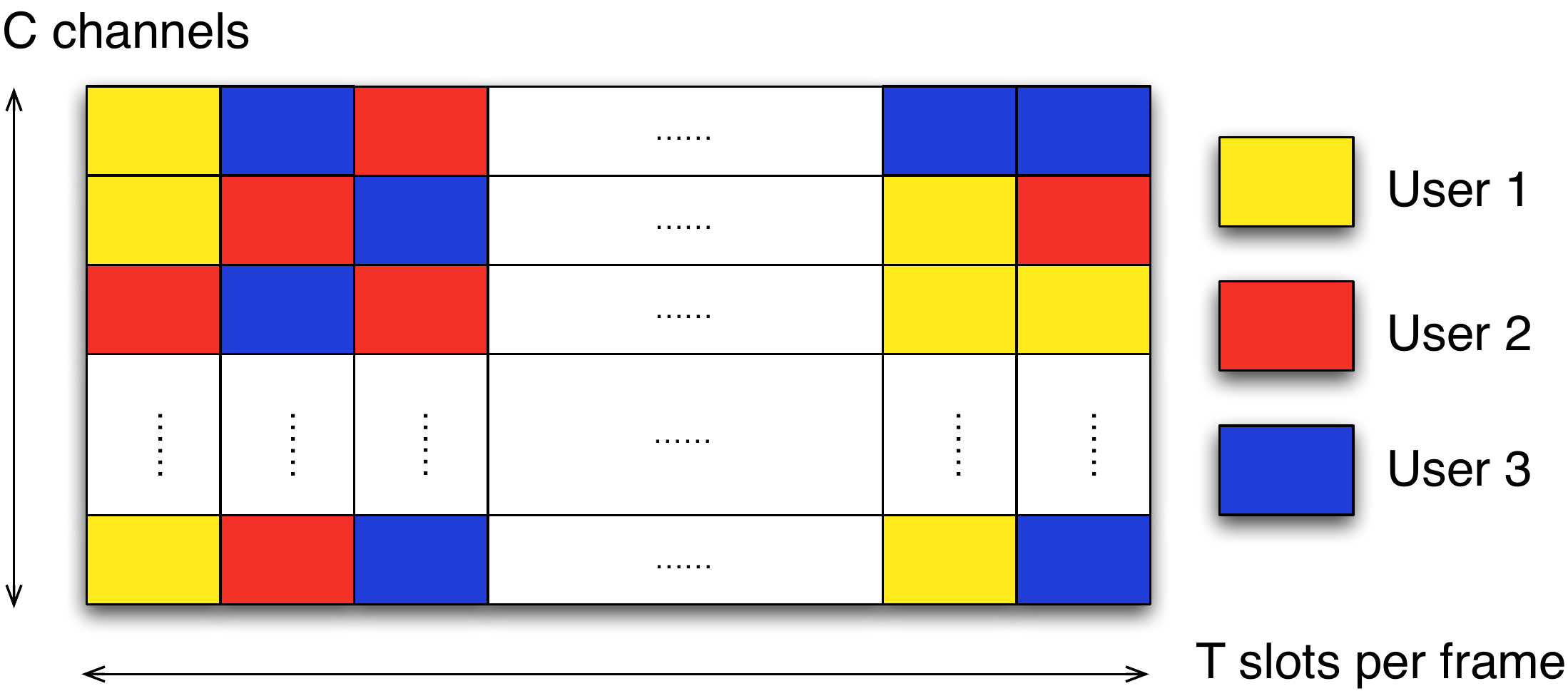}}
    \caption{
       A typical OFDMA frame in a temporal-frequency domain.
    }
    \label{fig:ofdma}
    \vspace{-0.1in}
\end{figure}
A typical OFDMA system partitions the radio resource in both frequency and time domains as shown in Figure~\ref{fig:ofdma}. The total bandwidth is divided into $C$ orthogonal subchannels, and one frame is divided into $T$ time slots. A subchannel and time slot combination, termed a \textit{tile}, is the minimum unit for resource allocation. In one BS, a tile can only be assigned to one user to avoid intra-cell interference. For each BS $b$, we use $\mathcal{M}_b$ to denote the set of users associated with it. Thus there are totally $|\mathcal{M}_b|$ users in this cell. For a specific BS, $\mathcal{M}_b$ may vary depending on the different cooperation strategies the operators use. Specifically, in NOCOOP, $\mathcal{M}_b$ are the users from the same operator of $b$ that are connected to $b$, whereas in FLEXROAM and MERGER, $\mathcal{M}_b$ include the users from the other operators that treat $b$ as the best BS. In addition, the number of subchannels $C$ may also vary according to the cooperation strategy. It will increase in MERGER as each BS reuses the spectrum owned by the other operators.

In the multi-cell scenario, neighboring cells may reuse the same tile depending on the inter-cell interference. As a result, the optimal multi-cell resource allocation problem is combinatoric and generally regarded as an NP-hard problem. Only heuristic methods exist so far~\cite{Koutsopoulos@TON06}\cite{Guoqing@TWC06}\cite{Honghai@JSAC11}. Next, we present a centralized greedy algorithm to achieve a sub-optimal solution with low computational complexity. It assumes the users' channel state information is available at a Radio Network Controller (RNC), which coordinates all the BSs. Note that since our purpose is to quantify the gain brought by the operator cooperation, designing more practical algorithms is beyond our scope. Our algorithm is generalized from the ones in~\cite{Koutsopoulos@TON06}\cite{Guoqing@TWC06}. It runs for each frame and aims to maximize the total utilities of all the users throughout all the cells in a frame. The utility of user $i$ is defined as $U(R_i) = \log(R_i)$, with $R_i$ being its data rate in the current frame. Using $\log$ function as the utility has been proved to be equivalent to finding a proportional fair solution~\cite{Kelly@Fair98}. Assuming there are $B$ BSs and $N$ users, the complexity of the algorithm is $O(TCBN)$.
\vspace{-0.1in}
\begin{algorithm}
\small {
$\mathbf{R}_{K \times T} = \mathbf{0}$; $\mathbf{Y}_{B \times C \times F} = \mathbf{0}$; $\mathbf{X}_{K \times C \times T} = \mathbf{0}$\;
\For{$t = 1 : T$} {
    \For{$c = 1 : C$} {
        $\Omega \leftarrow$ set of all the BSs\;
        \While{$\Omega \ne \emptyset$}{
        		$b^{*} \leftarrow$ BS with the least number of assigned channels in $\Omega$\;
        $\Omega \leftarrow \Omega \setminus b^{*}$\;
		$\mathbf{Y'} \leftarrow $ set $Y_{b^*, c, t} = 1$ in $\mathbf{Y}$\;
		\ForEach{$m \in \mathcal{M}_{b^{*}}$}{
%			$U_{gain} = \log(R_{m, f} + r_{m, c, f}(\mathbf{Y'}))$\
%			$ - \log(R_{m, f})$\;			
%			$U_{loss} = \sum_{i, X_{i, c, f} = 1} (\log(R_{i, f}) - \log(R_{i, f} - r_{m, c, f}(\mathbf{Y}) + r_{m, c, f}(\mathbf{Y'}))$\;
			$\Delta U_{m} = U_{gain, m} - U_{loss, m}$\;
		}
		$m^* \leftarrow \arg\max_{m} \Delta U_{m}$\;
		\If{$\Delta U_{m^*} > 0$} {
			$\mathbf{X} \leftarrow$ set $X_{m^{*}, c, t} = 1$ in $\mathbf{X}$\;
			$\mathbf{R} \leftarrow$ $\forall i, X_{i, c, t} = 1$, set $R_{i, t} = R_{i, t} - r_{i, c, t}(\mathbf{Y'}) + r_{i, c, t}(\mathbf{Y})$ in $\mathbf{R}$\;
			$\mathbf{Y} \leftarrow \mathbf{Y'}$\;
		}
        }
    }
}
}
\label{alg:1}
\caption{Multi-Cell Resource Allocation Algorithm}
\end{algorithm}
\vspace{-0.1in}

As shown in Algorithm 1, Line 1 initializes the matrices that contain the allocation results. Each entry $R_{i, t}$ in $\mathbf{R}$ records the aggregate data rate of user $i$ up to the slot $t$. $\mathbf{X}$ and $\mathbf{Y}$ are boolean matrices. The entry $Y_{b, c, t} = 1$ in $\mathbf{Y}$ if BS $b$ is transmitting on subchannel $c$ in slot $t$. The entry $X_{i, c, t} = 1$ in $\mathbf{X}$ indicates that the user $i$ is receiving on subchannel $c$ in slot $t$. The subchannels are allocated sequentially (Line 3), and for each one, we check each BS for whether it should transmit on this subchannel in the current slot. To maintain fairness, the BSs with fewer number of assigned subchannels are given higher priority (Line 6-7). Our algorithm is marginal utility driven. For the current BS $b^{*}$, Line 9-13 picks the user with the maximal marginal utility $\Delta U$. For each user $m$ associated with $b^*$, $\Delta U_m$ is defined as the difference between the utility increase $U_{gain, m}$ by scheduling user $m$ and the total utility loss of other users on the same subchannel $U_{loss, m}$ due to the increased interference (Line 10). $U_{gain, m}$ and $U_{loss, m}$ can be evaluated by:
\begin{equation}
U_{gain, m} = U[R_{m, t} + r_{m, c, t}(\mathbf{Y'})] - U(R_{m, t})
\end{equation}
\begin{equation}
U_{loss, m} =  \sum_{i, X_{i, c, t} = 1} \{U(R_{i, t}) - U[R_{i, t} - r_{m, c, t}(\mathbf{Y}) + r_{m, c, t}(\mathbf{Y'})]\}
\end{equation}
$\mathbf{Y'}$ is the matrix assuming BS $b^*$ is transmitting on subchannel $c$ in slot $t$. The function $r_{m, c, t}(\mathbf{Y})$ is the rate of user $m$ on subchannel $c$ in slot $t$ based on the allocation matrix $\mathbf{Y}$. We assume the transmit power is equally split across the subchannels on each BS. Considering the inter-cell interference, $r_{m, c, t}(\mathbf{Y})$ is defined as:
\begin{equation}
r_{m, c, t}(\mathbf{Y}) = W_C \log_2(1 + \frac{P_{t} / C \cdot h_{b_0, m}^c d_{b_0, m}^{-\alpha}}{N_0 W_C + \sum_{b \ne b_0, Y_{b, c, t} = 1}h_{b, m}^{c} d_{b, m}^{-\alpha}}),
\end{equation}
where $W_C = W/C$ is the bandwidth of a subchannel, $b_0$ is the BS user $m$ associates to. $h_{b, m}^{c}$ is the subchannel fading between BS $b$ and user $m$ on subchannel $c$, and $d_{b, m}$ is the distance between BS $b$ and user $m$. Finally, Line 12-15 update the matrices if the maximal marginal utility is positive.

\section{Performance Evaluation}

In this section, we assess the performance of our proposed cooperation strategies through two sets of simulations. In the first set, we numerically compute the expressions given by the analytical model described in Sec. II. To further show the results under practical settings, in the second set, we conduct simulations for the OFDMA system with the multi-cell resource allocation algorithm shown in Sec. III.

\begin{figure}[htbp]
    \centerline{\includegraphics[scale=0.20]{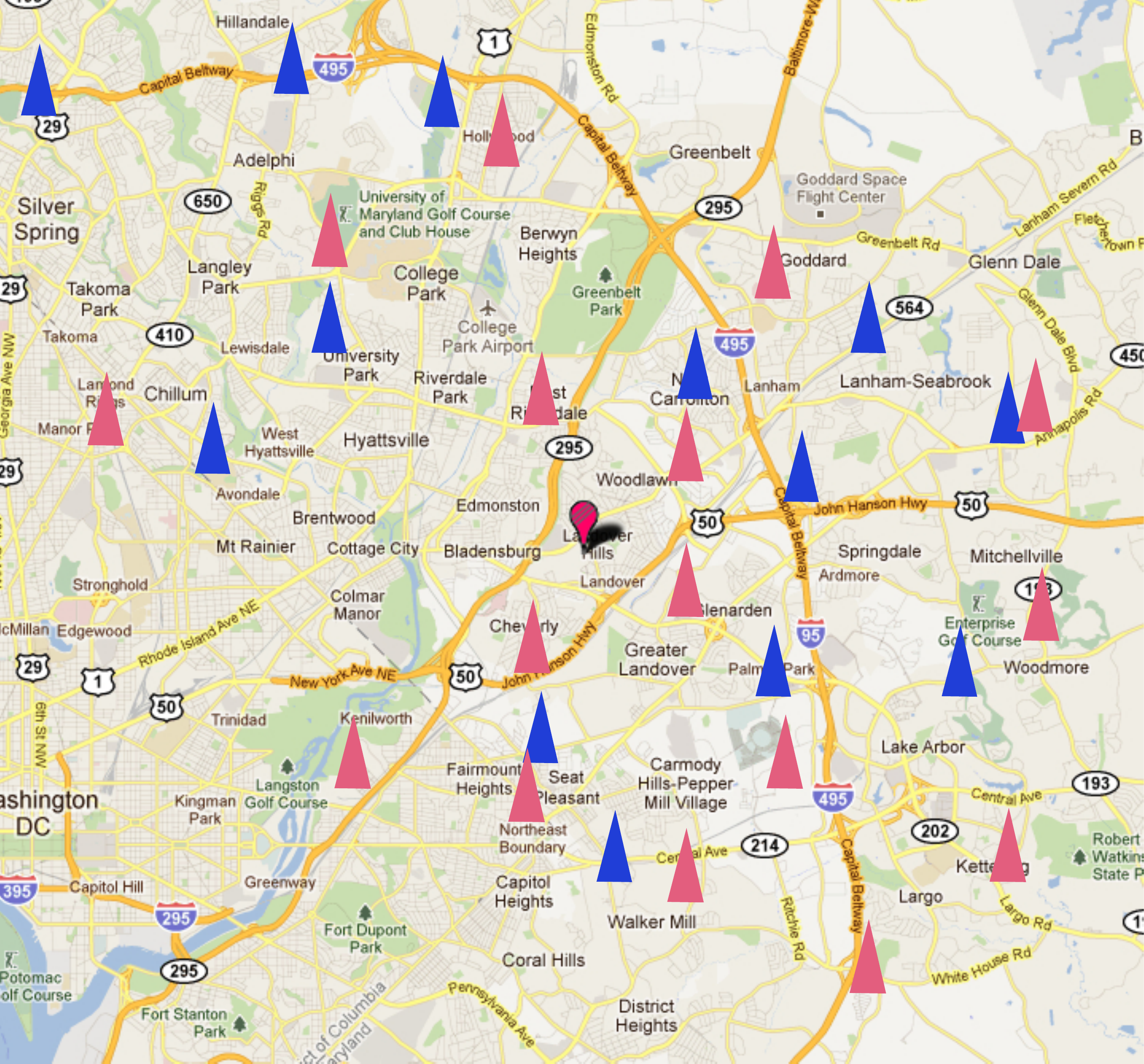}}
    \caption{
       A $20 \times 20$ km view of the current BS deployment by two major cellular operators in the Washington D.C. area.
    }
    \label{fig:dctowers}
\vspace{-0.3in}
\end{figure}

We have obtained precise coordinates for BSs from two major operators over a large suburban area near Washington D.C.. The $20 \times 20$ km area we chose is shown in Figure~\ref{fig:dctowers}. There are 16 BSs from one operator and 13 BSs from the other. We mark the BSs using red and blue to distinguish them. Their location information will be used for our simulation.

\subsection{Numerical Evaluations}

Here we numerically compute the results of our analytical model when different cooperation strategies are employed by the operators. We will focus on user's average throughput as it is a critical metric for evaluating cellular service.  It is important to show how much improvement there is through operator cooperation, and how such an improvement varies when the two operators have different BS densities, user densities and bandwidths. We set the value of $\lambda_1$ in Eq.~(12)-(14) as the red operator's BS density, e.g., $\lambda_1=16/400000000=4*10^{-8}$. $\lambda_1$ will be fixed in all the scenarios while $\lambda_2$ is left as an adjustable parameter. According to IEEE 802.16m evaluation methodology document~\cite{WiMAXEva}, we set the transmission power of the BSs as $Pt=46$ dBm, noise power density as $N_0 = -174$ dBm/Hz and the path loss exponent $\alpha = 3.76$.
%\begin{figure*}[htbp]
%    \centering
%    \subfigure[]
%    {
%        \includegraphics[scale=0.42]{changelambda.pdf}
%        \label{fig:changelambda}
%    }
%    \hspace{-0.35in}
%    \subfigure[]
%    {
%        \includegraphics[scale=0.42]{changeeta.pdf}
%        \label{fig:changeeta}
%    }
%    \hspace{-0.35in}
%    \subfigure[]
%    {
%        \includegraphics[scale=0.42]{changeW.pdf}
%        \label{fig:changeW}
%    }
%    \caption{Numerical results: (a) The impact of BS density. (b) The impact of user density. (c) The impact of bandwidth.}
%    \label{fig:numerical}
%\end{figure*}
\begin{figure*}[htbp]
\begin{minipage}[b]{0.36\linewidth}
\centering
\includegraphics[scale=0.42]{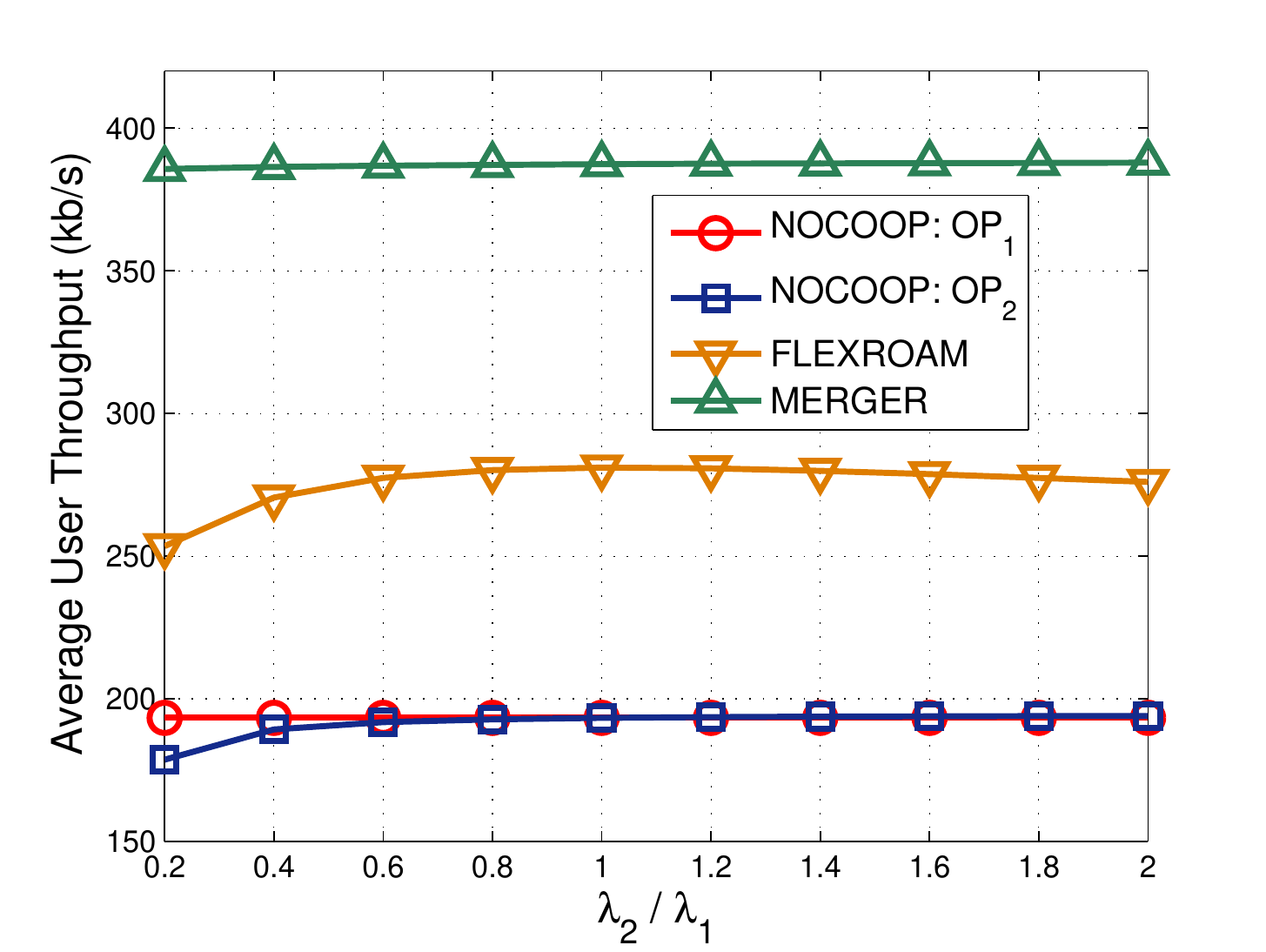}
\caption{The impact of BS density.}
\label{fig:changelambda}
\end{minipage}
\hspace{-0.37in}
\begin{minipage}[b]{0.36\linewidth}
\centering
\includegraphics[scale=0.42]{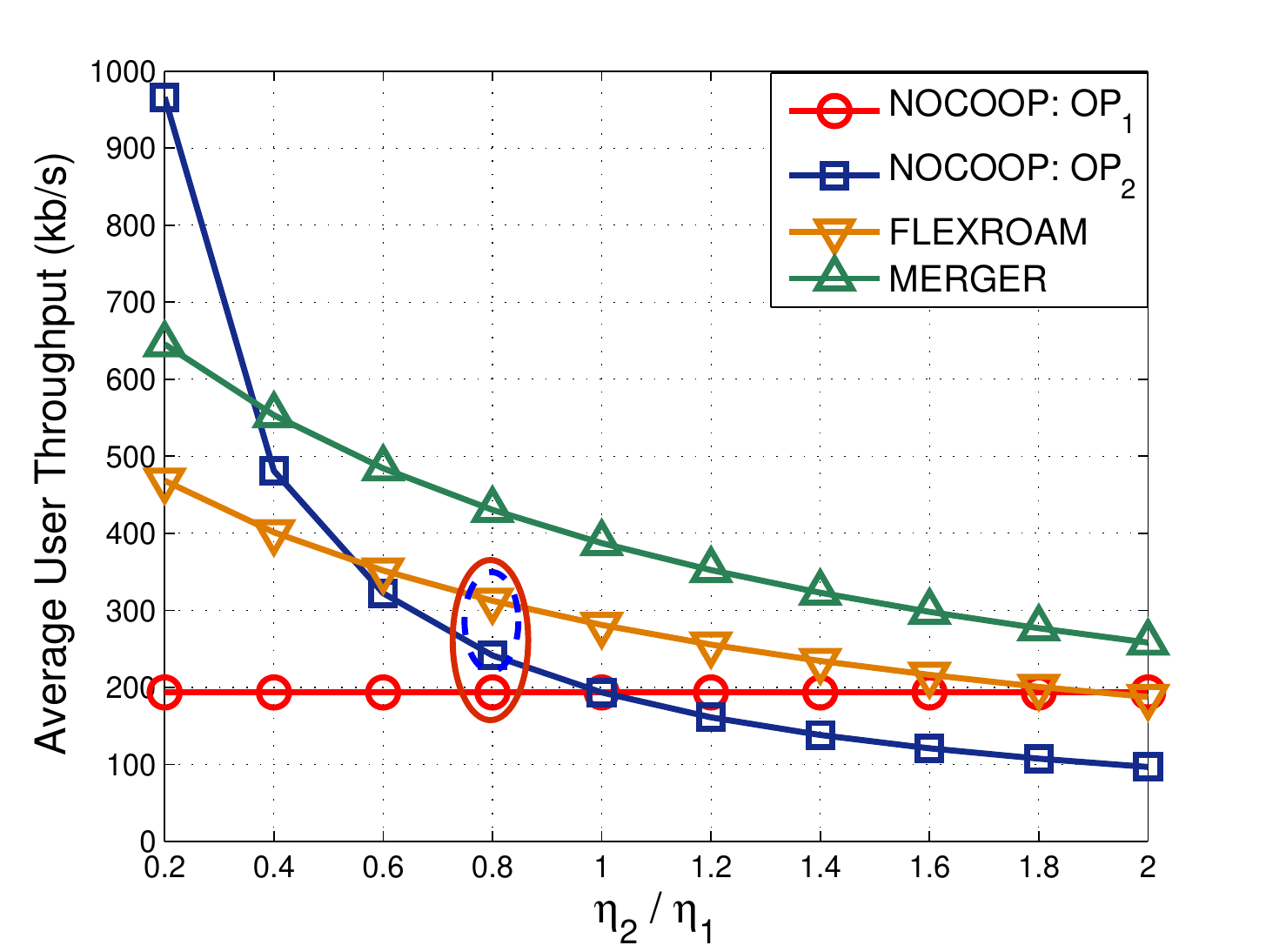}
\caption{The impact of user density.}
\label{fig:changeeta}
\end{minipage}
\hspace{-0.37in}
\begin{minipage}[b]{0.36\linewidth}
\centering
\includegraphics[scale=0.42]{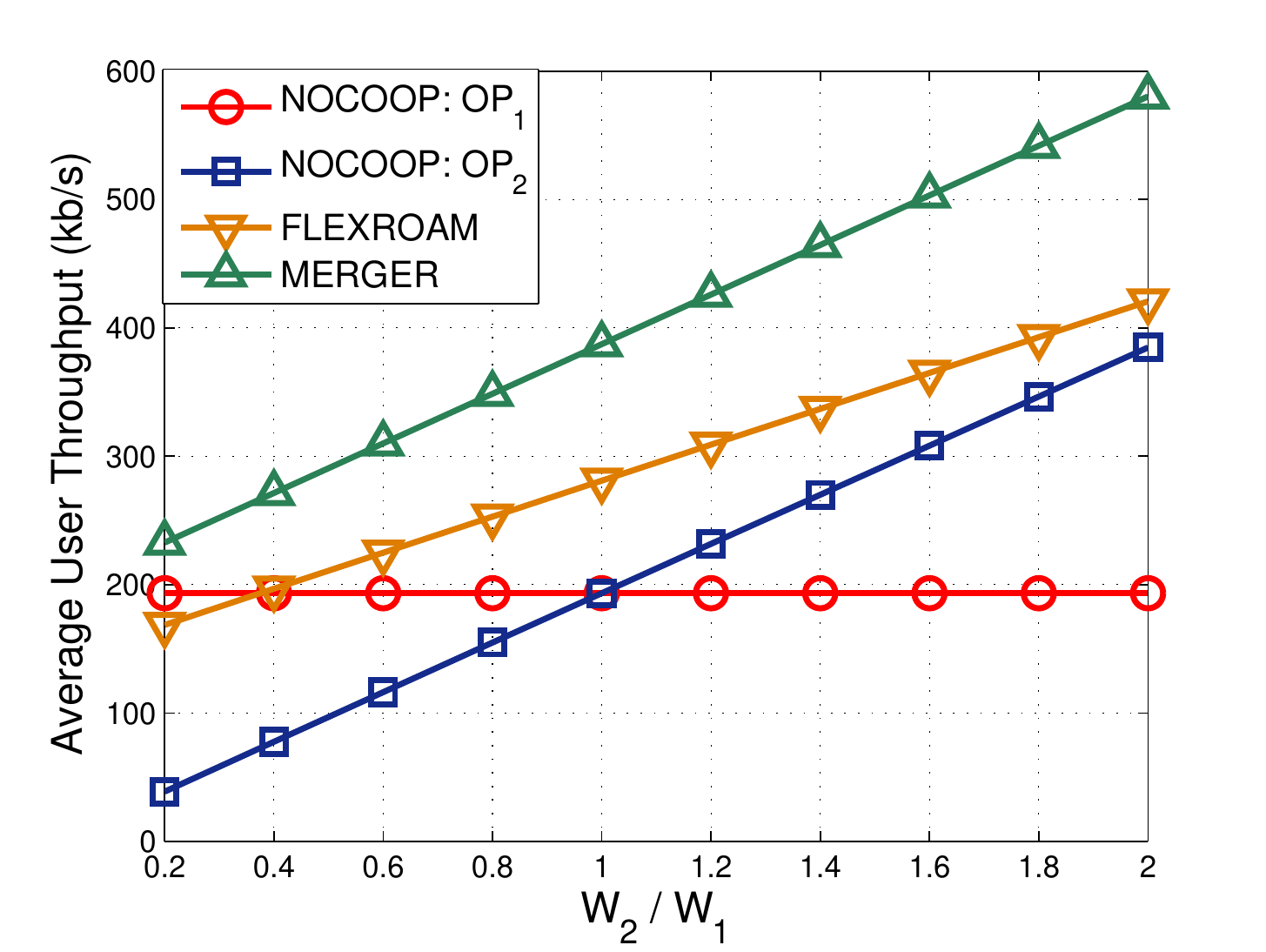}
\caption{The impact of bandwidth.}
\label{fig:changeW}
\end{minipage}
%\vspace{-0.3in}
\end{figure*}

\subsubsection{The Impact of BS Density}

In this scenario, we set both OP$_1$ and OP$_2$'s bandwidths to $W_1 = W_2 = 10$ MHz according to~\cite{WiMAXEva}. And we let their respective user densities be $\frac{\eta_1}{\lambda_1} = \frac{\eta_2}{\lambda_2} = 100$. As a result, their spectrum resource and average number of users per cell are identical. Figure~\ref{fig:changelambda} shows the results by changing the BS density ratio $\lambda_2 / \lambda_1$. As $\lambda_2$ changes from $0.2 \lambda_1$ to $2 \lambda_1$, 	the performance of OP$_2$ under NOCOOP only slightly increases, since $\lambda$ does not influence the signal-to-interference ratio and the system is interference limited. We can easily see cooperation brings significant benefits to both operators. When $\frac{\lambda_2}{\lambda_1} = 1$, the two operators have 193.3 kb/s average user throughput without cooperation. However, if they cooperate, FLEXROAM and MERGER strategies improve the average user throughput to 281.0 kb/s and 387.4 kb/s respectively, which is equivalent to a 45.4\% and 100.4\% increase, respectively.  As $\lambda_2$ keeps increasing, the performance of FLEXROAM will slightly drop since the opportunistic diversity is reduced. However, FLEXROAM and MERGER still achieve a large performance gain for both operators within a wide range of BS density ratios.
%\begin{figure}[htbp]
%    \centerline{\includegraphics[scale=0.55]{changelambda.pdf}}
%    \caption{
%       The impact of BS densities of the two operators.
%    }
%    \label{fig:changelambda}
%\end{figure}
%\begin{figure}[htbp]
%    \centerline{\includegraphics[scale=0.55]{changeeta.pdf}}
%    \caption{
%       The impact of user densities of the two operators.
%    }
%    \label{fig:changeeta}
%\end{figure}

\subsubsection{The Impact of User Density}

To explore the impact of user densities on the network performance of cooperation, we fix $\lambda_2 = \lambda_1 = 4*10^{-8}$ and $W_2 = W_1=10$ MHz. Then we let $\eta_1 = 100 \lambda_1$ and vary $\eta_2$ from 0.2$\eta_1$ to 2$\eta_1$. This represents different user/BS ratio for OP$_2$, e.g., when $\eta_2 = 0.2\eta_1=20\lambda_2$, on average there are only 20 users per cell.  Figure~\ref{fig:changeeta} shows the results. As we can see, as $\eta_2$ increases, under NOCOOP, the average user throughput of OP$_2$ drops drastically when more users are sharing the radio resource. In contrast, under FLEXROAM and MERGER the performance degrades gradually since users of OP$_2$ can still opportunistically connect to the BSs of OP$_1$. Moreover, when OP$_1$ and OP$_2$'s user densities are different, they achieve different performance gains. For instance when $\frac{\eta_2}{\eta_1} = 0.8$, under NOCOOP, OP$_1$ and OP$_2$ have 193.26 kb/s and 241.58 kb/s average user throughput. When FLEXROAM is adopted, the performance jumps to 312.21 kb/s, which gives 61.5\% and 29.2\% gains to OP$_1$ and OP$_2$, respectively (marked in the circles in the figure). Generally, FLEXROAM results in a win-win situation when $\frac{\eta_2}{\eta_1}$ ranges from around 0.5 to 2.0. MERGER has a larger win-win range, which is from 0.37 to beyond 2.0. When the difference between $\eta_1$ and $\eta_2$ is too large, cooperation may generate negative effects to one operator. For example, when $\frac{\eta_2}{\eta_1} = 0.4$, OP$_2$ loses performance under FLEXROAM. However, since we just are studying a simple roaming policy, such negative effects are expected to be avoided by adopting more advanced policies and incentive mechanisms.

%\begin{figure}[htbp]
%    \centerline{\includegraphics[scale=0.55]{changeW.pdf}}
%    \caption{
%       The impact of bandwidths of the two operators.
%    }
%    \label{fig:changeW}
%\end{figure}
\subsubsection{The Impact of BS Bandwidth}

In this scenario, we fix $\eta_1 = \eta_2 = 100\lambda_1 = 100\lambda_2$. We fix $W_1 = 10$ MHz, and change the ratio $W_2 / W_1$. As we can see from Figure~\ref{fig:changeW}, the performance of FLEXROAM, MERGER and OP$_2$ under NOCOOP increases almost linearly as the total bandwidth increases. When $\frac{W_2}{W_1} > 0.4$, FLEXROAM provides better performance for both OP$_1$ and OP$_2$. It enables more gain for the operator with less spectrum. MERGER achieves a larger gain for all the points in the figure.

\subsection{Performance with Real BS Locations}

To validate the numerical results in practice, we conducted the OFDMA-based simulations with real BS locations. We consider two cellular networks with BSs placed as in Figure~\ref{fig:dctowers}. Mobile devices are deployed uniformly in the 20 km $\times$ 20 km area with a density of 100 users per cell for each operator. The system is assumed to be coordinated by the OFDMA resource allocation algorithm we proposed in Sec. III. Based on the IEEE 802.16m evaluation methodology document~\cite{WiMAXEva}, the main system parameters are summarized in Table~\ref{table:ofdmapara}. For the channel gain, we model the fast fading component as Rayleigh fading with mean 1. We also considered shadow fading and model it as a log-normal random variable with a standard deviation of 8 dB. The overall effect of antenna gain, cable and penetration loss and noise figure is set to 0 dB. Finally, the path loss at the distance $d$ is modeled as $L(d)_{dB} = 17.39 + 3.76\log_{10}d$.
\begin{table}[!t]
\caption{OFDMA System Parameters}
\label{table:ofdmapara}
\centering
\begin{tabular}{ p{5.5cm} | l }
\hline
Number of subchannels & 32 \\
\hline
Number of slots per frame & 60 \\
\hline
BS Transmit Power & 46 dBm \\
\hline
Noise power spectrum density $N_0$ & -174 dBm \\
\hline
Channel bandwidth & 10 MHz \\
\hline
\end{tabular}
\vspace{-0.2in}
\end{table}

Depending on the different cooperation strategies, the user set of each BS may be different. Under NOCOOP and FLEXROAM, each operator still aims to optimize the total utility of all the users connected to its own BSs, while under MERGER, we optimize the global utility of the two operators. We ran each simulation for 30 consecutive frames and we averaged the results of 5 simulation runs. Further, we plug the density values $\lambda_1 = 16 / 400000000 = 4*10^{-8}$, $\lambda_2 = 13 / 400000000=3.25*10^{-8}$, $\eta_1 = 100 \lambda_1$, $\eta_2 = 100 \lambda_2$ into our previous analytical model to resolve the numerical results for comparison purpose.

\begin{figure}[htbp]
    \centerline{\includegraphics[scale=0.55]{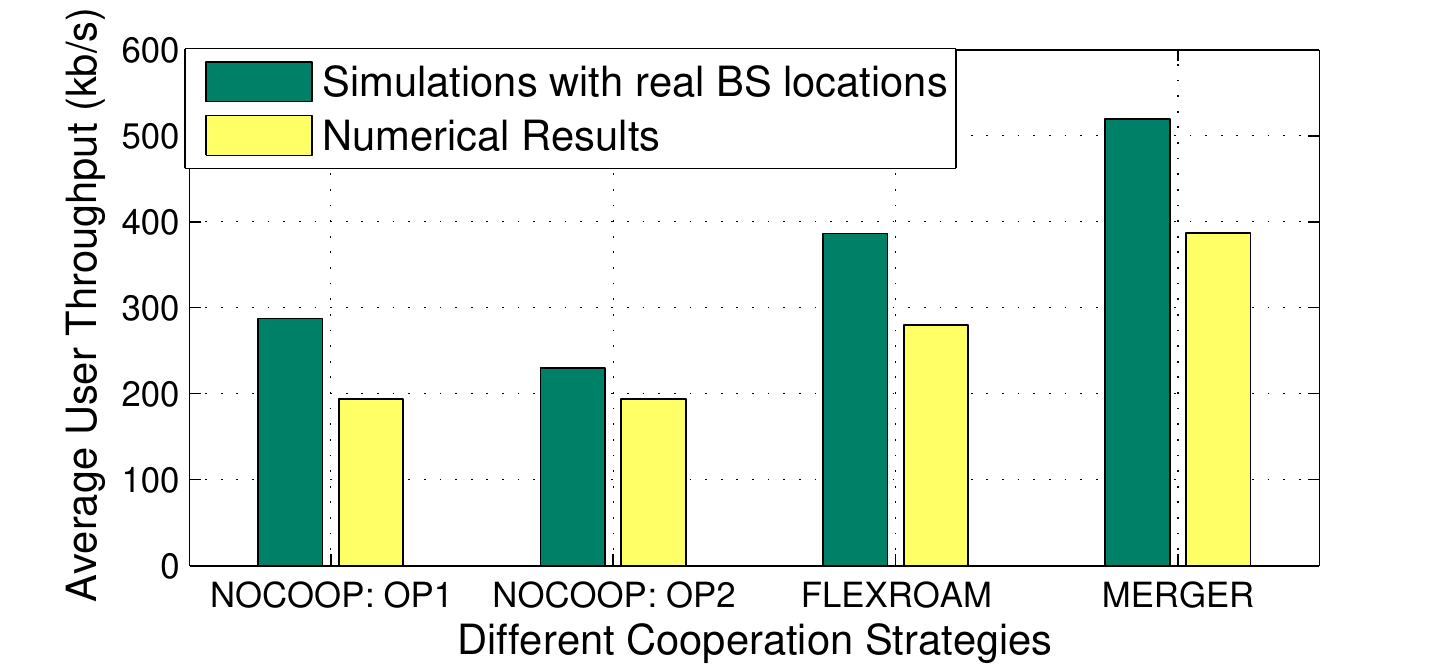}}
    \caption{
       network performance under OFDMA-based system.
    }
    \label{fig:ofdmabar}
    \vspace{-0.1in}
\end{figure}
Figure~\ref{fig:ofdmabar} shows the results. Under NOCOOP, the performance of OP$_2$ is worse than OP$_1$ since it has fewer BSs and the resulting average signal is weaker. Overall, the average throughput of all the users in the system under NOCOOP is around 263.8 kb/s. Cooperation substantially improves the performance of both the operators the real setting. Specifically, the user's average throughput increases to 386.24 kb/s in FLEXROAM and 519.31 kb/s in MERGER. These give 34.4\% and 80.75\% improvement to OP$_1$ and 68.21\% and 126.16\% improvement to OP$_2$ respectively. Moreover, our analytical model accurately captures this trend and provides very close improvement predictions in terms of percentage. OFDMA system generally offers better performance than the numerical results since it better exploits the user/frequency selectivity. Also, the analytical model is pessimistic since it assumes the random deployment of BSs, which is discussed in~\cite{Jeffrey@TCOM10}.

\begin{figure}[htbp]
    \centerline{\includegraphics[scale=0.55]{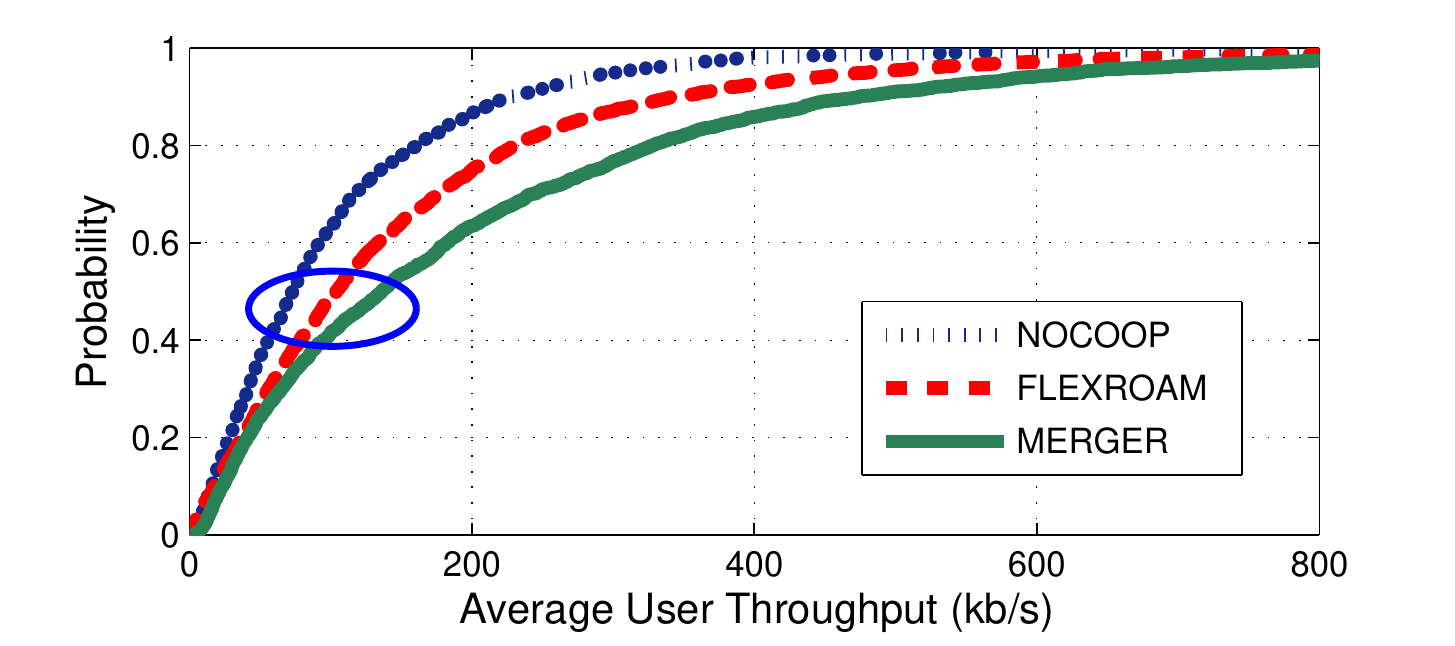}}
    \caption{
       user's average throughput CDF.
    }
    \label{fig:cdf}
    \vspace{-0.2in}
\end{figure}
To further scrutinize the gain for the users, we draw the CDF of the average throughput of the users in one simulation run. Figure~\ref{fig:cdf} shows the results in the domain of 0-200 kb/s. For NOCOOP, we use the statistics from all the users in the system. The  curves validates the improvement result from cooperation. For example, the median value of user's average throughput increases by around 40\% and 100\% under FLEXROAM and MERGER compared to NOCOOP.

\section{Conclusion}

This work investigates the potential benefits of cooperation among cellular operators. Using stochastic geometry, we provide a tractable analytical model to derive the average user rate and throughput under two cooperation strategies. With real base station locations, extensive simulations over multi-cell OFDMA system further validate the performance improvement. We show even a simple cooperation policy with modest changes to existing networks can achieve around 30\%-120\% capacity gains per customer under typical conditions.

\bibliographystyle{IEEEtran}

\end{document}